\documentclass[aps,prx,twocolumn,superscriptaddress,groupedaddress]{revtex4-1}
\usepackage{graphicx,amsmath,color}

\def\w{\omega}

\def\w{\omega}

\def\bk{{\bf k}}
\def\bK{{\bf K}}

\def\bG{{\bf G}}

\def\bl{{\bm l}}
\def\bL{{\bm L}}

\def\e{\epsilon}
\def\ve{\epsilon}

\def\<{\langle}
\def\>{\rangle}

\usepackage{amsfonts}
\usepackage{braket} % for Dirac notation bras and kets

 \usepackage{tikz}
\def\t{\text}
\usepackage{bm}
\usepackage{ulem}
\renewcommand{\vec}[1]{\bm{#1}}

\begin{document}

\title{Fully Anharmonic, Non-Perturbative Theory of\\
Vibronically Renormalized Electronic Band Structures}

\author{Marios Zacharias}
\author{Matthias Scheffler}
\author{Christian Carbogno}
\affiliation{Fritz Haber Institute, Theory Department, Faradayweg 4-6, 14195 Berlin Germany}

\date{\today}

\begin{abstract}
We develop a first-principles approach for the treatment of vibronic interactions in solids that 
overcomes the main limitations of state-of-the-art electron-phonon coupling formalisms. In particular,
anharmonic effects in the nuclear dynamics are accounted to all orders via {\it ab initio} molecular 
dynamics simulations. This non-perturbative, self-consistent approach evaluates the response of 
the wave functions along the computed anharmonic trajectory; thus it fully considers the coupling 
between nuclear and electronic degrees of freedom. We validate and demonstrate the merits of the 
concept by calculating temperature-dependent, momentum-resolved spectral functions for silicon and the 
cubic perovskite SrTiO$_3$, a strongly anharmonic material featuring soft modes. In the latter case, 
our approach reveals that anharmonicity and higher-order vibronic couplings contribute substantially 
to the electronic-structure at finite-temperatures, noticeably affecting band gaps and effective masses,
and hence macroscopic properties such as transport coefficients.
%demonstrated here for the electron effective masses.
%noticeably affecting  macroscopic properties, 
%such as absorption coefficients as well as thermal and electrical conductivities. 
%\MZ{Our present approach opens the way for the correct description of the properties of 
%materials suitable for high-temperature applications}
\end{abstract}

%\pacs{
%71.15.-m, %      Methods of electronic structure calculations
%71.38.-k  %      Polarons and electron-phonon interactions
%}
\maketitle

\section{Introduction} \label{Intro}

Electronic band structures are a fundamental concept in material science used to 
qualitatively understand and quantitatively assess optical and electronic properties 
of materials,~e.g.,~charge carrier mobilities and absorption spectra of semiconductors. 
Over the last decade, three pivotal advancements have paved the way towards predictive, 
quantitative \textit{ab initio} calculations of electronic band structures: advances in relativistic
approaches~\cite{Blaha:2020ci}, improvements in 
the treatment of electronic exchange and correlation~\cite{Golze:2019fx,Chen:2012cf} and the 
inclusion of electron-phonon interactions via perturbative many-body formalisms based on the 
Allen-Heine theory~\cite{Allen_Heine_1976}. The latter approach has been widely used to calculate 
temperature-dependent effects on the electronic structure stemming from the nuclear 
motion~\cite{Marini_2008,Giustino_2010,Cannuccia_2011,Cannuccia_2012,Antonius_2014,Ponce_2014,Ponce_2014_2,Kawai_2014,Ponce_2015, Molina_2016,Zhou_2016,Menendez_2017,Ponce_2018,Querales_2019,Park_2020}.
However, such perturbative calculations rely on {\it\textbf{two}} approximations. 
a) The nuclear motion is approximated in a harmonic model which is equivalent to the concept of phonons and 
b) the vibronic interaction between electronic and nuclear degrees
of freedom is treated by perturbation theory in terms of electron-phonon coupling. 
In both approximations, interactions at finite temperatures~$T$ 
are thus described via truncated Taylor expansions, using derivatives computed at the static equilibrium 
geometry,~i.e.,~for the total energy minimum corresponding to the atomic geometry~$\vec{R}^\t{eq}$ obtained 
in the classical $T=0$~K limit. Clearly, both these approximations are problematic whenever large deviations 
from~$\vec{R}^\t{eq}$ occur,~e.g.,~at elevated temperatures and for soft bonded atoms. 
Several strategies have been proposed to mitigate either one of these 
approximations~\cite{Monserrat2013,Patrick_2015,Antonius_2015,Monserrat_2014,Zacharias_2015,Zacharias_2016,Bartomeu_2016,Bartomeu_2016_GW,Shulumba_2016}. This revealed that the predictive power of perturbative 
calculations can be problematically limited, even for low-temperature properties of  
simple materials such as MgO and LiF~\cite{Antonius_2015}; solids featuring more anharmonic
dynamics, such as molecular crystals~\cite{Monserrat:2015jj}, and perovskites~\cite{Saidi_2016} which are affected more severely. A consistent computational approach able to settle these issues 
by accounting on equal footing for {\it \textbf{both}} anharmonic effects in the nuclear motion 
and the full vibronic coupling is, however, still lacking.

In this work, we fill this gap by deriving a fully anharmonic, non-perturbative first-principles 
theory of vibronic coupling and demonstrate its implementation in the all-electron, 
numeric atomic orbitals code \textit{FHI-aims}~\cite{Blum_2009}. As a validation, we show that our approach 
reproduces literature data for Si, a largely harmonic case in which the perturbative approach
works exceptionally well. Furthermore, we compute temperature-dependent spectral functions, band gaps, and
effective masses for cubic SrTiO$_3$, a prototypical perovskite. In this case, 
the highly-anharmonic dynamics~\cite{Rupprecht_1961,Cowley_1962} associated to the octahedral-tilting typically observed in 
perovskites~\cite{Woodward:1997ci,Lee:2016bz} results in a breakdown of the perturbative %electron-phonon coupling 
model and thus in significant changes of the electronic properties.
%and in a significant reduction of the band gap as well as an enhancement of the effective masses. 
Besides clarifying the experimental findings for SrTiO$_3$~\cite{Kok_2015,Allen_S_2103},
%reconciling previous discrepancies between theory and experiment for SrTiO$_3$~\cite{Allen_S_2103}, 
our calculations reveal that anharmonic, higher-order vibronic couplings~(AVICs) 
have substantial influence on the electronic properties, especially of perovskites, a material class with exceptional 
potential for high-temperature applications~\cite{Schultz:2015hi,Marina:2002kka,Skinner:2001fv,Ohta:2007fua,Fergus:2012jha,Brunauer:2016jv}.
%For all these applications, 
%an accurate assessment of temperature-dependent, momentum-resolved electronic band structures and derived 
%properties is essential.
%=======
%calculations thus suggest that accounting for anharmonic, higher-order vibronic coupling~(AViC) is essential in perovskites, 
%a material class with exceptional application potential in high-temperature thermoelectric waste-heat recovery devices~\cite{Fitriani_2016} and in hybrid photovoltaic cells operated under concentrated sunlight~\CC{CITE}. To
%advance both these fields, an accurate assessment of temperature-dependent, momentum-resolved electronic 
%structure is essential, so that fundamental material properties such as band gaps and effective masses~\cite{Huo:2018cs,Pei:2012ih}, but also the anisotropic band-structure corrugation~\cite{Chen:2013cn} can be reliably determined 
%at all.
%>>>>>>> 48e2ef269425d2b824c8f657d0d1b5cb634854a9

The organization of the manuscript is as follows: in Sec.~\ref{Theory}
we introduce the theoretical framework of our statistically anharmonic, higher-order vibronic coupling ({\it stAVIC}) 
approach for obtaining full anharmonic temperature-dependent band structures. In the same section we also 
outline the main equations involved in our implementation of band structure unfolding using numeric atom-centered orbitals.
In Sec.~\ref{sec.Methods} we present all computational details of the calculations
performed in this work. 
In Sec.~\ref{sec.Results} we demonstrate the merits of our methodology by 
reporting first principles calculations of temperature-dependent spectral functions and band gaps 
of Si and SrTiO$_3$ for a wide range of temperatures. 
For SrTiO$_3$, we also report temperature-dependent effective masses. 
Section~\ref{Sec.Conclusions} summarizes our key results and emphasizes the 
importance of our methodology for materials' design in various applications. 
Further technical details are left to Appendices~\ref{app.thermal_exp}-\ref{app.JDOS}.

\section{Theory} \label{Theory}
In this section we describe the theoretical framework of our methodology and discuss the implementation details 
of the band structure unfolding technique when numeric atom-centered orbitals are used.

\subsection{ Statistically anharmonic, higher-order vibronic coupling ({\it stAVIC}) approach}

In the following, the energy $\ve^{\vec{R}}_{l}$ of the electronic state $\ket{\psi^{\vec{R}}_{l}}$ is obtained 
by solving the Schr\"odinger equation $H_{ \rm el}^{\vec{R}} \ket{\psi^{\vec{R}}_l} = \ve^{\vec{R}}_{l} \ket{\psi^{\vec{R}}_l}$, 
where $H_{ \rm el}^{\vec{R}}$ is the electronic Hamiltonian of the system at the atomic 
geometry ${\vec{R}}$. This may be a Kohn-Sham Hamiltonian with a certain exchange-correlation functional. 
For readability, we use the generalized index $l$ to indicate both the band index~$n$ and the 
wave vector~$\vec{k}$. The temperature dependence of $\ve^{\vec{R}}_{l}$ is evaluated within 
the Born-Oppenheimer approximation via the canonical ensemble average at temperature $T$:
\begin{equation}\label{eq.energies_T}
 \<\ve^{\vec{R}}_{l}\>_T = \frac{1}{Z}\int d\vec{R}d\vec{P} \exp{\bigg[\frac{-E(\vec{R},\vec{P})}{k_{\rm B}T}\bigg]} \ve^{\vec{R}}_{l} .
\end{equation}
Here, $k_{\rm B}$ is the Boltzmann constant, $Z$ %=\int d\vec{R}d\vec{P} \exp[-E(\vec{R},\vec{P})/k_{\rm B}T]$
the canonical partition function, $\vec{P}$ the momenta of the nuclei, 
and $E(\vec{R},\vec{P})$ the total energy of the combined electronic and nuclear system. 
For the evaluation of Eq.~\eqref{eq.energies_T}, the state-of-the-art formalism~\cite{FG_review} resorts 
to the {\it \textbf{two}} perturbative approximations mentioned above. When the harmonic approximation to the potential 
energy surface (PES) is employed, the classical equations of motions for~$\vec{R},\vec{P}$ can be solved 
analytically, and so can the quantum-mechanical Schr\"odinger equation. 
Hence, Eq.~\eqref{eq.energies_T} is approximated via $\<\ve^{\vec{R}}_{l}\>_T \approx \<\ve^{\vec{R}}_{l}\>_T^\t{ha}$ as
\begin{equation}\label{eq.energies_ha}
\<\ve^{\vec{R}}_{l}\>_T^\t{ha} = \frac{1}{Z^\t{ha}} \int d\vec{R}^\t{ha}d\vec{P}^\t{ha} \exp{\bigg[\frac{-E^\t{ha}(\vec{R}^\t{ha},\vec{P}^\t{ha})}{k_{\rm B}T}\bigg]} \ve^{\vec{R}}_{l} \;,
\end{equation}
which allows for a straightforward evaluation of the phase-space 
integral~\cite{Zacharias_2015,Zacharias_2016,Bartomeu_2016,Monserrat_2018,Zacharias_2020}. 
When the dependence of the electronic states on 
the nuclear motion is truncated up to second order in the atomic 
displacements~$\ve^{\vec{R}}_{l} \approx \ve^{{{\rm pt},\vec{R}}}_{l}$, then the ensemble average 
in Eq.~\eqref{eq.energies_ha} yields the perturbative Allen-Heine energies $\<\ve^{{\rm pt},{\vec{R}}}_{l}\>_T^\t{ha}$.

In this work, we rely on neither of the two approximations. %.discussed above. 
First, \textit{ab initio} molecular dynamics~(aiMD) trajectories with length~$t_0$ are used to 
evaluate the canonical ensemble average in Eq.~(\ref{eq.energies_T}) as time~($t$) average 
\begin{eqnarray} \label{eq.aiMD_T}
   \<\ve^{\vec{R}}_{l}\>_T = \<\ve^{\vec{R}}_{l}\>^{\rm MD}_T = 
\lim_{t_0 \rightarrow \infty} \frac{1}{t_0}\int_0^{t_0}  \ve^{{\vec{R}}(t)}_{l} dt \;. 
\end{eqnarray}
This accounts for the full anharmonicity of the PES. Second, the dependence of the 
electronic eigenenergies $\ve_l^{\vec{R}(t)}$ on the nuclear positions
is explicitly evaluated by solving 
$H_{ \rm el}^{\vec{R}(t)} \ket{\psi^{\vec{R}(t)}_l} = \ve^{\vec{R}(t)}_{l} \ket{\psi^{\vec{R}(t)}_l}$ 
at each aiMD step~$\vec{R}(t)$. All orders of coupling between electronic and nuclear degrees of 
freedom are included by these means. This involves re-expanding
\begin{equation}
\label{expansion}
\ket{\psi^{\vec{R}(t)}_l}=\sum_m p_{ml}^{\vec{R}(t)}  \ket{\psi^\t{eq}_m} 
 \text{ with } p_{ml}^{\vec{R}(t)} = \braket{\psi^\t{eq}_m|\psi^{\vec{R}(t)}_l} 
\end{equation}
in terms of the wave functions at equilibrium~$\ket{\psi^\t{eq}_m}$.
With that, one obtains:
\begin{eqnarray} \label{eq:eps_of_t}
 \epsilon_l^{\vec{R}(t)} & = & \bra{\psi^{\vec{R}(t)}_l} H_{\rm el}^{\vec{R}(t)} \ket{\psi^{\vec{R}(t)}_l}
                            = \epsilon_l^\t{eq} + \\
&  & \sum_{m,n}  [p_{nl}^{\vec{R}(t)}]^*p_{ml}^{\vec{R}(t)}   \bra{\psi^\t{eq}_n} 
H_{\rm el}^{\vec{R}(t)}- H_{\rm el}^{\t{eq}}\ket{\psi^\t{eq}_m} \;.  \nonumber
\end{eqnarray}
In this form, it is evident that Eq.~\eqref{eq:eps_of_t} not only incorporates the first 
non-vanishing derivatives of~$H_{\rm el}^{\vec{R}(t)}- H_{\rm el}^{\t{eq}}$ as perturbative formalisms, but 
all orders. Similarly, all orders of couplings with the nuclear motion -- not just quadratic terms -- are 
captured via the coefficients $p_{ml}^{\vec{R}(t)}$, which describe the intricate $\vec{R}(t)$-dependence 
of the wave functions along the aiMD. Accordingly, all orders of AVIC are {\it statistically} 
captured by these means. Our approach, named \textit{stAVIC} in the 
following, is thus valid even when the (harmonic) phonon ansatz is inappropriate. 

\begin{figure*}[htb]
\includegraphics[width=0.75\textwidth]{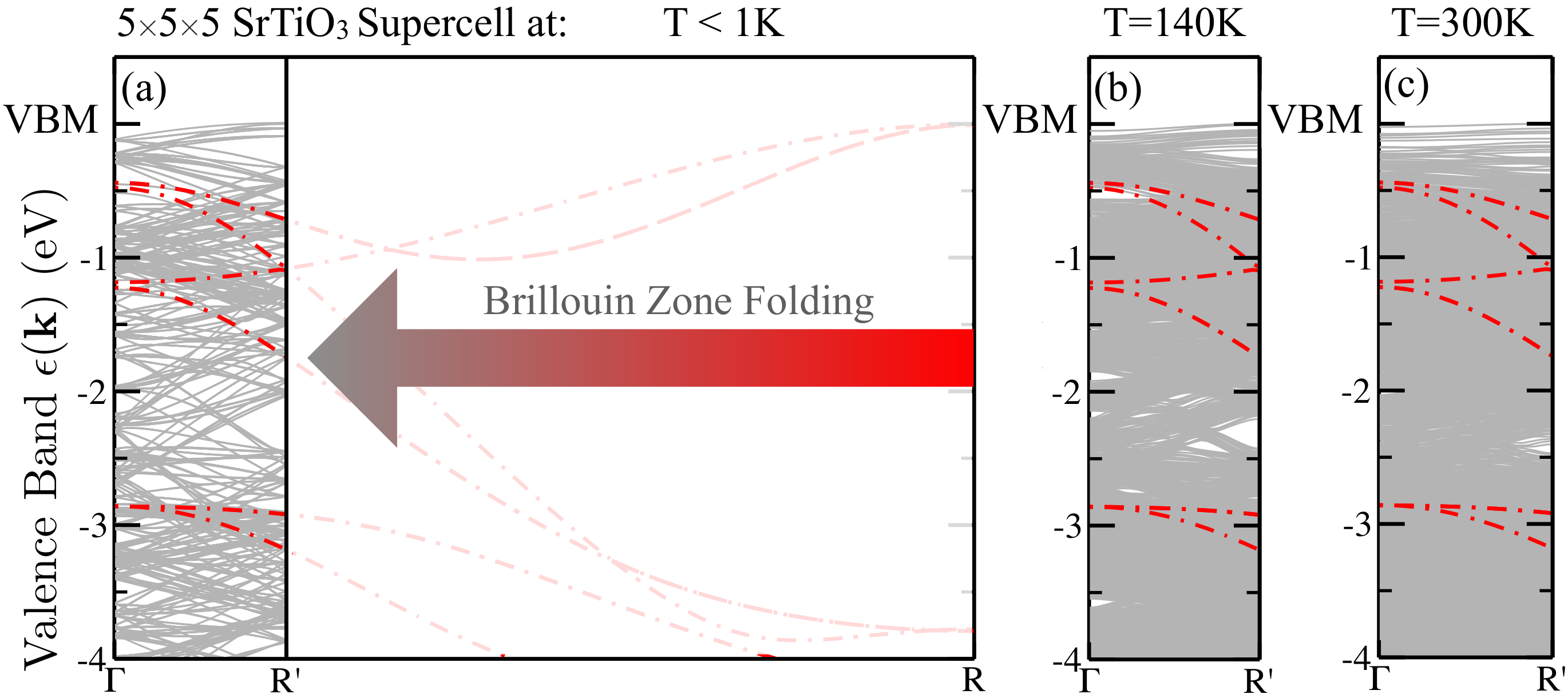}
  \caption{\label{fig1} Electronic valence band structure of SrTiO$_3$ along $\Gamma$-R obtained 
from calculations in the primitive unit cell with static nuclei at equilibrium (red) and in 5$\times$5$\times$5 supercells (grey). 
As plot~(a) exemplifies, the electronic dispersion~$\epsilon(\vec{k})$ obtained in the primitive unit cell along 
$\Gamma$-R is ``folded'' into a reduced Brillouin zone with a shorter reciprocal-space path $\Gamma$-R' in supercell calculations. 
While individual states and their momentum-dependence is still visible in case of tiny displacements 
(plot a, $T<1$K), this is no longer the case at finite temperatures, since the nuclear motion breaks the symmetries within the supercell. 
Accordingly, momentum-resolved electronic structures are no longer accessible, as plot (b) and (c) show 
for representative geometries obtained from aiMD runs at 140 K and 300K, respectively.}
\end{figure*}

In practice, the thermodynamic average in Eq.~(\ref{eq.aiMD_T}) can be evaluated via 
\textit{ab initio} path-integral MD~\cite{Ramirez:2006bg}, or via aiMD~\cite{Franceschetti:2007dg}, 
the latter corresponding to the classical, high-temperature limit of interest in this work. 
Regardless, a direct evaluation 
of Eq.~(\ref{eq.aiMD_T}) is not particularly useful, since it does not give access to state- and 
momentum-resolved band-structures in the fundamental Brillouin zone~(BZ).
Since large supercells are required to capture vibrations with non-zero wavevector in solids, 
the obtained electronic energies~$\epsilon_l=\epsilon_{N\vec{K}}^{\vec{R}(t)}$ and wave 
functions~$\psi_l = \psi_{N\vec{K}}^{\vec{R}(t)}$, with band indices $N$ and wave vectors~$\vec{K}$, only
span a reduced BZ~\cite{Boykin:2005bh} (capital letters indicate supercell quantities). As shown in Sec.~\ref{SWF}, individual 
states thus become indistinguishable and only band edges can be reliably identified~\cite{Ramirez:2006bg,Franceschetti:2007dg}. 
Besides preventing a comparison with ARPES experiments or with the static limit at~$\vec{R}^\t{eq}$, 
for which the wave vectors~$\vec{k}$ of $\epsilon_l^\t{eq}=\epsilon_{n\vec{k}}^\t{eq}$ 
and $\psi_l^\t{eq} = \psi_{n\vec{k}}^\t{eq}$ span the full fundamental BZ,
this ``BZ folding'' makes it impossible to determine state- and momentum-dependent electronic properties, such
as lifetimes and effective masses. To recover a band 
structure in the fundamental BZ also for supercells, the expansion coefficients introduced 
in Eq.~\eqref{expansion} are used to ``unfold'' the states~$\psi_{N\vec{K}}$.
To this aim, we consider the spectral function expressed in the Lehman representation~\cite{Allen_2013}:
\begin{eqnarray}\label{eq.spectral_fn_diag}
A^{\vec{R}(t)}_{n \bk} (E) = \sum_{N \bK} |p_{n\vec{k},N\vec{K}}^{\vec{R}(t)}|^2 \delta(E - \ve^{\vec{R}(t)}_{N \bK}).
\end{eqnarray}
Compared to Eqs.~\eqref{expansion}-\eqref{eq:eps_of_t}, in which the perturbed 
eigenvalue~$\ve^{\vec{R}(t)}_{N \bK}$ is obtained from a superposition of equilibrium 
states~$\psi^{\rm eq}_{n \bk}$, Eq.~\eqref{eq.spectral_fn_diag} reflects the inverse relationship: 
Each perturbed eigenvalue~$\ve^{{\vec{R}}(t)}_{N \bK}$ contributes to all states~$n\vec{k}$ in the 
fundamental BZ, whereby~$p_{n\vec{k},N\vec{K}}^{\vec{R}(t)}= \braket{\psi^{\rm eq}_{n \bk}| \psi^{ {\vec{R}}(t)}_{N \bK} }$ 
determines the strength of this contribution. 

For each configuration~$\vec{R}(t)$, we obtain the momentum-resolved spectral 
function~$A^{{\vec{R}}(t)}_{\bk} (E)= \sum_n A^{{\vec{R}}(t)}_{n \bk} (E)$  by summing 
over~$n$ in  Eq.~\eqref{eq.spectral_fn_diag}:
\begin{eqnarray}\label{eq.spectral_fn_diag_unres2}
A^{{\vec{R}}(t)}_{\bk} (E) = \sum_{N \bK} P^{{\vec{R}}(t)}_{\bk, N \bK} \delta(E - \ve^{{\vec{R}}(t)}_{N \bK}) \;.   
\end{eqnarray}
The spectral weight~$P^{{\vec{R}}(t)}_{\bk, N \bK} = \sum_n |p_{n\vec{k},N\vec{K}}^{{\vec{R}}(t)}|^2$ 
describes the overlap between the supercell state $\ket{\psi^{{\vec{R}}(t)}_{N\bK}}$ and all 
equilibrium states with wave vector $\bk$~\cite{Popescu_2012,Medeiros_2014}. 
Details for its numerical evaluation are provided in Sec.~\ref{SWF}.
The momentum-resolved spectral function in thermodynamic equilibrium~$\<  A^{{\vec{R}}(t)}_{\bk} (E) \>_T$ is 
then computed as the thermodynamic average of $A^{{\vec{R}}(t)}_{\bk}(E)$ along~$\vec{R}(t)$ 
via Eq.~\eqref{eq.aiMD_T}. Momentum-resolved quasi-particle peaks are extracted 
from~$\< A^{{\vec{R}}(t)}_{\bk} (E) \>_T$ by scanning over the energy axis, 
from which band gaps~$\<\ve_{\t{g}}\>_T$ and effective masses~$m_e^*$ are then obtained.

\subsection{stAVIC: Spectral Weight Formalism} \label{SWF} 

In \textit{ab initio} MD simulations of solids, it is necessary to use extended supercells to 
accurately sample vibrations with non-zero wavevector. Due to the larger cell size in real space, 
the electronic structure obtained in reciprocal space from such supercell calculation suffers from BZ folding.
In other words, the band structure is associated to a reduced BZ, as shown in Fig.~\ref{fig1}(a). 
At finite temperatures, the motion of the nuclei breaks the symmetries within the supercell, thus making individual states indistiguishable, 
cf.~Fig.~\ref{fig1}(b) and (c). This prevents any assessment of the momentum-dependence of the electronic dispersion. Hence, a BZ 
unfolding~\cite{Boykin:2005bh} is necessary to reverse this BZ folding, so to obtain clearly disentangled states in the fundamental BZ, 
and so to map the properties of the dynamical system back onto the established language and terms of solid-state physics.

%\MZ{As a sanity check, we first apply our implementation for the calculation of the spectral function of unperturbed supercell configurations, i.e. using classical nuclei and $T=0$~K. Figure~\ref{figS0}(a) shows the band structure (colored plot) of silicon calculated for an unperturbed 6$\times$6$\times$6 supercell configuration before unfolding in comparison to the unperturbed band structure calculated using the unit cell (black line). Figure~\ref{figS0}(b) shows the spectral function (colored plot) of silicon calculated for an unperturbed 6$\times$6$\times$6 supercell configuration after unfolding [using Eq.~\eqref{eq.spectral_fn_diag_unres}] in comparison to the unperturbed band structure calculated using the unit cell (black line).  As expected, the two plots overlay perfectly. For completeness we also present in Figs.~\ref{figS0}(c) and ~\ref{figS0}(d) the corresponding band structure and spectral function of Si calculated using the stAVIC approach at $T = 500$~K. The differences between the spectral functions in Figs.~\ref{figS0}(b) and~\ref{figS0}(d) reflect the effect of vibronic coupling on the band structure.}

\begin{figure*}[hbt!] 
  \includegraphics[width=0.98\textwidth]{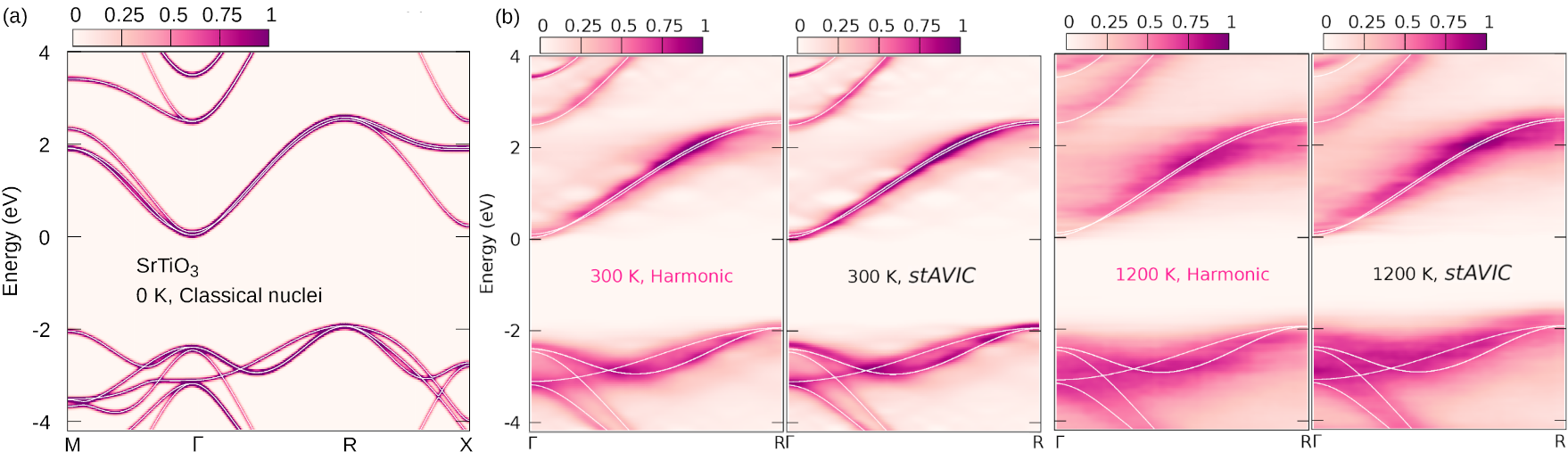}
  \caption{  
  \label{fig2}
(a) Spectral function $A_{\bk} (E)$ of cubic SrTiO$_3$ along the high-symmetry path M-$\Gamma$-R-X (336 $\bk$-points) calculated 
using DFT-PBE in a $5\times5\times5$ supercell containing 625 atoms at their classical, 0K positions (static equilibrium).
(b) Thermodynamically averaged spectral functions $\<  A^{{\vec{R}}(t)}_{\bk} (E) \>_T$ of cubic SrTiO$_3$ along 
$\Gamma$-R (128 $\bk$-points) for $T=300$~K and $1200$~K calculated non-perturbatively on the harmonic 
and anharmonic PES using DFT-PBE and 30 uncorrelated geometries in a $5\times5\times5$ supercell 
containing 625 atoms. For all plots the band structure in static equilibrium is shown as white lines. }
\end{figure*}

The band resolved spectral function $A_{n \bk} (E)$ can be obtained from the 
imaginary part of the retarded one-electron Green's function~\cite{Bassani_book} 
and then evaluated using Eq.~\eqref{eq.spectral_fn_diag}.
In practice, the electronic states are expanded~\cite{Blum_2009} as a linear combination of Bloch-type
functions using the expansion coefficients $c_{ j,n\bk}$ and $C_{ J,N\bK}$, respectively,
\begin{eqnarray}
\psi_{n \bk} = \sum_j c_{ j,n\bk}\; \chi_{j,\bk} \text{  and  }\psi_{N \bK} = \sum_J C_{ J,N\bK}\; \chi_{J,\bK}  \,.
\end{eqnarray}
For notational clarity we have dropped the superscript indices eq and ${\bf R}(t)$  
from the reference $\psi_{n \bk}$ and perturbed $\psi_{N \bK}$ states, respectively.
The Bloch-type functions are related to atomic orbitals via discrete Fourier transforms
\begin{eqnarray}
\chi_{j,\bk} & = & \sum_{\bl} e^{-i\bk \cdot \bl} {\phi_{j,\bl}} \\
\chi_{J,\bK} & = & \sum_{\bL} e^{-i\bK \cdot \bL} {\Phi_{J,\bL}} \;,
\end{eqnarray}
in which the sums are taken over all lattice vectors $\bl = (l_1,l_2,l_3)$ and  $\bL = (L_1,L_2,L_3)$.
Enforcing translational invariance yields the overlap matrix elements $p_{n\vec{k},N\vec{K}}$ 
in the following form~\cite{Ku_2010,Lee_2013}:
\begin{eqnarray}\label{eq.spectral_matel}
p_{n\vec{k},N\vec{K}} & = &\braket{ \psi_{n \bk}|\psi_{N \bK}} \\ &=& \sqrt{\frac{L}{l}}
\sum_{jJ} c^*_{ j,n\bk} C_{ J,N\bK} \nonumber \\
&\times& \sum_{\bl} e^{-i\bk \cdot \bl}  
\braket{\phi_{j,\bl}| \Phi_{J,{\bf 0}} } \delta_{\bk - \bG, \bK}, \nonumber
\end{eqnarray}
Here, the first summation runs over all 
real basis functions $\ket{\phi_{i,\bl}}$ and $\ket{\Phi_{J,{\bf 0}}}$ of the reference and perturbed system, respectively. 
%The corresponding expansion coefficients are denoted by $c_{ j,n\bk}$ and $C_{ J,N\bK}$. 
%The second summation runs over all translation lattice vectors $\bt(\bl)[\, \bl = (l_1,l_2,l_3) \,]$ 
The corresponding Born-von Karman supercells 
contain $l$ and $L$
periodic replicas of the original cell, set at ${\bf 0}$, along each Cartesian direction. %, with $l=l_1 \times l_2 \times l_3$. 
The presence of the Kronecker delta $\delta_{\bk - \bG, \bK}$ ensures that $\bK$ is mapped onto 
$\bk$ via a reciprocal lattice vector $\bG$ of the perturbed system. 
For the mapping between the indices of the basis functions of the reference 
and perturbed system, $j$ and $J$, the following relationship is satisfied: 
\begin{eqnarray}\label{eq.mapping_mat}
\sum_{j'}  s^{-1}_{jj'}(\bk)S_{j'J}(\bK) = \delta_{jJ}
\end{eqnarray}
with
\begin{eqnarray}
 s_{jj'}(\bk) &=& \sum_{\bl} e^{i\bk \cdot \bl} \braket{\phi_{j,{\bf 0}}| \phi_{j',\bl}}, \,\, \text{and} \\
\label{eq.s_1_tran}
 S_{j'J}(\bK) &=& \sqrt{\frac{L}{l}}\sum_{\bl} e^{-i\bk \cdot \bl} \braket{\phi_{j',\bl} | \Phi^0_{J,{\bf 0}}} \delta_{\bk - \bG, \bK}.
\label{eq.s_2_tran}
\end{eqnarray}
Here $\ket{\Phi^0_{J,{\bf 0}}}$ indicates the basis functions of an unperturbed supercell, in which the 
atomic nuclei are at their equilibrium positions.
%the system that has the 
%same dimensions as the perturbed system but is an exact replica of the reference system. 
In other words, $\ket{\Phi^0_{J,{\bf 0}}}$ are the equilibrium states in the supercell, 
obtained by periodically replicating the unperturbed reference system.
%,~i.e.,~in which no atomic displacements occur.
Taking the summation over all bands $n$ in Eq.~\eqref{eq.spectral_matel} yields  
the spectral weights $P_{\bk,N \bK}$ entering Eq.~\eqref{eq.spectral_fn_diag_unres2}. 
In particular, using the completeness 
relation~$\mathbb{I}_\bk= \sum_n \ket{\psi_{n \bk}}\bra{\psi_{n \bk}}$ of the 
states  $\ket{\psi_{n\bk}}$ the spectral weight can be re-written as:
\begin{eqnarray}\label{eq.spectral_weight}
P_{\bk, N \bK} =  \braket{ \Psi_{N \bK} | \mathbb{I}_\bk | \Psi_{N \bK}} , 
\end{eqnarray}
and the final result with respect to the perturbed expansion coefficients 
and overlap matrix of the perturbed basis functions is: 
\begin{eqnarray}\label{eq.spectral_weight_2}
P_{\bk, N \bK} &=&  \frac{L}{l}
\sum_{jJ \bl}  C^*_{ J,N\bK} C_{ j,N\bK}
 e^{-i\bk \cdot \bl}  
\braket{\Phi_{j,\bl}| \Phi_{J,{\bf 0}} }  \delta_{\bk - \bG, \bK}. \nonumber 
\\ 
\end{eqnarray}
The advantage of the above expression for the calculation of the spectral weights 
is that no knowledge of the wavefunctions of the reference system is explicitly required. 
% and (ii) the evaluation of the overlap matrix of the basis functions does not suffer from 
% the quadrature weight derivatives problem~\cite{Baker_1994}~(\textit{moving grid effect}) 
% when atom-centered orbitals are employed. 
% COMMENT Chris: Moving grid type effects can/should still be there for other matrix elements

In our implementation we evaluate Eq.~\eqref{eq.spectral_fn_diag_unres2}
 by calculating the spectral weights using Eq.~\eqref{eq.spectral_weight_2} 
and ensuring that the correct mapping between $j$ and $J$ indices is obtained 
through Eq.~\eqref{eq.mapping_mat}. 
In Fig.~\ref{fig2}(a) we demonstrate the first step of the validation of our approach by showing
 the perfect band structure unfolding as calculated for a $5\times5\times5$ supercell geometry of 
SrTiO$_3$, which is an exact periodic replica of the unit cell containing atoms at their relaxed classical positions.
Examples of spectral functions obtained by supercell calculations 
on perturbed configurations exploring the harmonic~(left) and anharmonic~(right) PES for $T=300$~K and $1200$~K 
are shown in Fig.~\ref{fig2}(b).
The relevant computational details are provided in Sec.~\ref{sec.Methods}.

\section{Computational details} \label{sec.Methods}

All calculations were performed with the all-electron, 
full-potential, numeric-atomic orbital code \textit{FHI-aims}~\cite{Blum_2009} 
using DFT-LDA for Si and DFT-PBE for SrTiO$_3$. In the latter case, van-der-Waals interactions 
were included using the Tkatchenko-Scheffler method~\cite{Tkatchenko_2009}.
For both structures, light defaults were used for the numerical settings and for the basis set. %whereas
With respect to Brillouin zone~(BZ) integrations, 
12$\times$12$\times$12~(Si) / 5$\times$5$\times$5~(SrTiO$_3$) $\vec{k}$-grids 
(in the primitive BZ) were used during the self-concistency cycle.

Table~\ref{table.1} summarizes the lattice constants and band gaps computed 
for static, cubic SrTiO$_3$ using different exchange-correlation 
functionals (LDA, PBE, PBEsol, and HSE06), with and without van der 
Waals interactions. %The latter were accounted for via the Tkatchenko-Scheffler formalism~\cite{Tkatchenko_2009}. 
Generally, we observe that vdW interactions stabilize the cubic structure, curing the typical underbinding
observed with the PBE functional. All LDA and GGA functionals severely underestimate the experimental band
gap of 3.26~eV~\cite{Kok_2015} by at least 1~eV, whereas HSE06 and HSE06-vdW yield static band gaps of
 3.63 and 3.45~eV, respectively. The fact that PBE-vdW yields an excellent 
agreement with respect to experimental lattice expansion data, validates our choice of the functional used for all  
{\it stAVIC} calculations on SrTiO$_3$.
 
\begin{table}[hbt!]
\begin{center}
\begin{tabular}{| c | c | c | }
\hline\hline
XC-Functional      & Lattice Constant (\AA) & Bang gap (eV)   \\ \hline 
LDA                & 3.86  & 2.00  \\ \hline                                     
PBE                & 3.96  & 2.30 \\  \hline
PBE-vdw            & 3.90  & 2.09 \\  \hline
PBEsol             & 3.90  & 2.10 \\   \hline
PBEsol-vdw         & 3.87  & 2.00 \\  \hline
HSE06              & 3.91  & 3.63 \\   \hline
HSE06-vdW          & 3.85  & 3.45 \\    \hline
Expt.~(T=140~K)     & 3.90 ~\cite{Farrel_1964}   &  3.26~\cite{Kok_2015}  \\ \hline\hline
\end{tabular}
 \caption{Lattice constant and band gap of cubic SrTiO$_3$ calculated within density functional theory using the 
LDA, PBE, PBEsol, and HSE06 exchange-correlation (XC) functionals. Van der Waals (vdW) interactions are accounted 
for via the Tkatchenko-Scheffler method~\cite{Tkatchenko_2009}.}
\label{table.1}
\end{center}
 \end{table}

Harmonic phonon properties were calculated using 
finite differences as implemented in the software package {\it PHONOPY}~\cite{phonopy}. 
The computed phonon frequencies and normal mode coordinates were employed to 
sample the harmonic phase space via \textit{Importance Sampling Monte Carlo}~\cite{Zacharias_2015} 
so to evaluate the thermodynamic averages~$\< \cdot \>_T^{\rm ha-qm}$ and
$\< \cdot \>_T^{\rm ha-cl}$ using quantum-mechanical and classical statistics, respectively.   
Soft modes with imaginary phonon frequencies are ``frozen in'' in this harmonic approach 
($\sim$1\% of all modes in the employed SrTiO$_3$ supercells).

Born-Oppenheimer \textit{ab initio} molecular dynamics~(aiMD) simulations 
were carried out in the canonical ensemble~(NVT) using a time-step of 1~fs 
and the Bussi-Donadio-Parrinello thermostat~\cite{Bussi_2007}. Trajectories 
with a time length of 1.5 - 2.0~ps for Si and SrTiO$_3$ were used to 
thermalize/equilibrate the systems. Additional 1.0 - 2.0~ps for Si and SrTiO$_3$
of aiMD were simulated for the \textit{stAVIC} evaluation. 
Thermal lattice expansion was calculated by computing the thermodynamic 
average of the stress tensor observed during an aiMD trajectory; 
subsequently, the structure was re-optimized under external pressure, 
so to obtain temperature dependent geometries for which the stress 
tensor becomes negligible in thermodynamic average~\cite{Roekeghem_2016}. 
As shown in Appendix~\ref{app.thermal_exp}, a significant band gap opening is induced
when considering thermal lattice expansion in SrTiO$_3$.

The band gaps at finite temperatures are obtained by averaging over 
50~(Si) and 100~(SrTiO$_3$) configurations, which are selected from  the already equilibrated aiMD trajectory in steps of 0.01~ps.
6$\times$6$\times$6 supercells with 432 atoms~(Si) and 5$\times$5$\times$5 supercells with 625 atoms~(SrTiO$_3$) were used for the aiMD. Both, the chosen number of configurations for the thermodynamic averaging and the chosen supercell sizes, ensure a
convergence of the temperature-dependent band-gap within~$\ll \pm 5$\%. We note that, finite supercell size effects 
and symmetry breaking lead to the splitting of degenerate states,
as observed for the triply degenerate valence band maximum of Si before~\cite{Zacharias_2016,Karsai_2018}.
In this case, we determine the energy change of the band as the 
mean renormalization of all originally degenerate bands~\cite{Karsai_2018}.
For high temperatures ($T>500~K$) the large quasiparticle linewidths lead to a large spectral 
broadening making the quasiparticle peaks of the band edges hard to distinguish. As a second check, 
band gaps were additionally determined by analyzing the thermodynamically averaged joint density of states~\cite{Zacharias_2016}, 
as discussed in Appendix~\ref{app.JDOS}. For all investigated temperatures, the analysis of the joint density of 
states confirmed our \textit{stAVIC} calculations.
The temperature-dependent electron effective-masses of SrTiO$_3$ were extracted from the calculated 
momentum-resolved spectral functions by performing parabolic fits of these spectral functions in the 
proximity of the conduction band minimum~($\Gamma$) along the corresponding high-symmetry paths 
connecting $\Gamma-R$ and $\Gamma-M$. 

Eventually, let us note that the computational cost is dominated by the sampling of the phase space,~i.e.,~the
\textit{ab initio} MD in the \textit{stAVIC} calculations. The numerical effort to perform
the BZ unfolding and obtain spectral functions is comparable to a few self-concistency cycles.

\section{Results} \label{sec.Results}

\begin{figure*}[t]
  \includegraphics[width=0.77\textwidth]{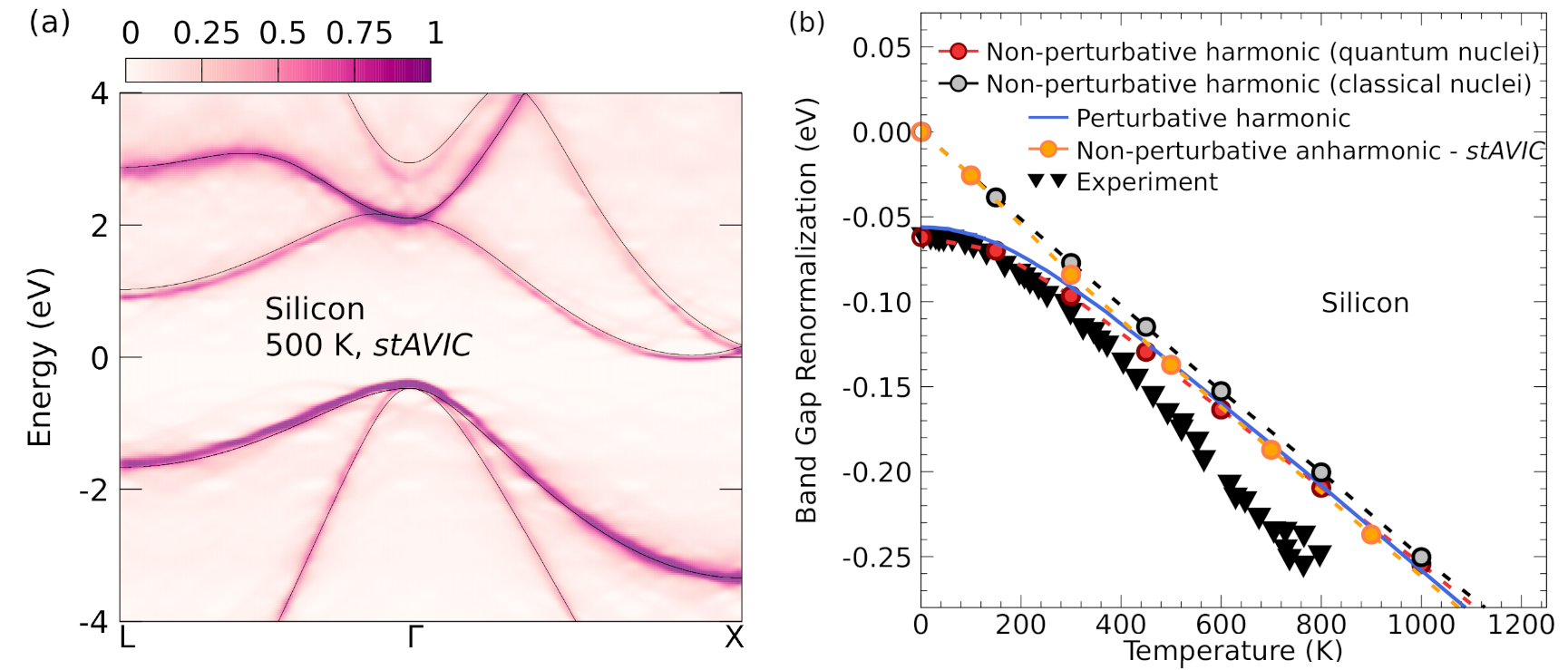}
  \caption{\label{fig3} 
   (a) Thermodynamically averaged spectral function $\<  A^{{\vec{R}}(t)}_{\bk} (E) \>_T$ of silicon 
   along L-$\Gamma$-X (256 $\bk$-points) for $T=500$~K calculated with \textit{stAVIC} using DFT-LDA and 30 uncorrelated geometries in
   a $6\times6\times6$ supercell containing 432 atoms. 
   The band structure in static equilibrium is shown as black lines.
   (b) Vibronic renormalization of the band 
   gap~$\Delta\<\epsilon_{\t{g}}\>_T = \<\epsilon_{\t{g}}\>_T-\<\epsilon_{\t{g}}\>_{0{\rm K}}^\t{ha-cl}$ of silicon 
   as function of temperature obtained via the \textit{stAVIC}~$\Delta\<\epsilon_{\t{g}}\>_T^\t{MD}$~(orange) 
   and via the non-perturbative harmonic 
   approach: $\Delta\<\epsilon_{\t{g}}\>_T^\t{ha-cl}$~(grey) and $\Delta\<\epsilon_{\t{g}}\>_T^\t{ha-qm}$~(red). 
   All calculations were performed using DFT-LDA and 6$\times$6$\times$6 supercells containing 432 atoms.
   Perturbative harmonic calculations~(blue, \cite{Ponce_2015})
   and experimental data~(black, \cite{Alex_1996}) are shown as well.
  }
\end{figure*}

In this section we demonstrate the potential of our methodology in calculating full temperature-dependent 
band structures and band gaps, that include anharmonic effects and all orders of vibronic coupling. 
As a validation, we show that our approach
reproduces harmonic data for Si, for which the perturbative Allen-Heine approach
performs particularly well, and then we present temperature-dependent spectral functions, band gaps, and
effective masses for cubic SrTiO$_3$.

\subsection{\textit{stAVIC}: Temperature-dependent band structure of Si}

Figure~\ref{fig3}(a) shows the momentum-resolved spectral-function of Si along the high-symmetry path L-$\Gamma$-X at 500~K,
as calculated using the \textit{stAVIC} approach. For comparison purposes we also include 
the band structure of Si calculated for the unit cell with the nuclei at static equilibrium.
The differences between the two plots reflect essentially the effect of vibronic coupling on the band structure.
For example, identifying the positions of quasiparticle peaks of the spectral function 
reveals that the valence band top at $\Gamma$ increases 
in energy by 73~meV and the conduction band bottom at 0.83 $\Gamma$-X lowers 
in energy by 64~meV leading to a total band gap renormalization of 137~meV.

Figure~\ref{fig3}(b) shows the temperature dependence of the 
band gap renormalization $\Delta \<\epsilon_{\t{g}}\>_T = \<\epsilon_{\t{g}}\>_T-\<\epsilon_{\t{g}}\>_{0{\rm K}}^\t{ha-cl}$ 
of bulk Si. Our aiMD-based \textit{stAVIC} calculations~$\Delta\langle\ve_{\t{g}}\rangle_T^\t{MD}$ 
are in excellent agreement with reference data~$\Delta\<\ve_{\t{g}}^\t{pt}\>_T^{\rm ha-qm}$ 
obtained with the perturbative, harmonic formalism~\cite{Ponce_2015} for $T\gg 400$~K. 
Discrepancies at lower temperatures are exclusively caused 
by quantum-nuclear effects not captured in aiMD. In Fig.~\ref{fig3}(b), this is demonstrated by 
comparing non-perturbative, harmonic data obtained by 
evaluating Eq.~\eqref{eq.energies_ha} with Monte Carlo 
sampling~\cite{Patrick2013,Zacharias_2015} using classical~$\Delta\<\ve_{\t{g}}\>_T^{\t{ha-cl}}$ 
and quantum-mechanical~$\Delta\<\ve_{\t{g}}\>_T^{\t{ha-qm}}$ statistics.
In both cases, anharmonic effects are thus neglected, while higher-order vibronic couplings 
are included via Eq.~\eqref{eq:eps_of_t}. The fact that the anharmonic $\Delta\<\ve_{\t{g}}\>_T^\t{MD}$ 
and the harmonic approach  $\Delta\<\ve_{\t{g}}\>_T^{\t{ha-cl}}$ almost coincide in the 
classical limit proves that anharmonic effects are indeed negligible for silicon and that
discrepancies with experiment at high~$T$ reflect the deficiencies of the LDA functional~\cite{Bartomeu_2016_GW,Karsai_2018}. 
Similarly, higher-order vibronic couplings are negligible here, given that the 
non-perturbative $\Delta\<\ve_{\t{g}}\>_T^{\t{ ha-qm}}$ and the perturbative 
data~$\Delta\<\ve_{\t{g}}^\t{pt}\>_T^{\t{ha-qm}}$ follow closely each other. 
Quantitatively, this is substantiated by the fact that 
our $\Delta\<\ve_{\t{g}}\>_T^{\t{ha-qm}}$~calculations yield a quantum zero-point renormalization 
of 62~meV in line with previous harmonic 
approaches~(56-62~meV)~\cite{Monserrat_2014,Bartomeu_2016,Zacharias_2015,Zacharias_2016} and 
with experimental values~(62-64~meV)~\cite{Cardona20053,Cardona_2001}.

\begin{figure*}[hbt!]
  \includegraphics[width=0.46\textwidth]{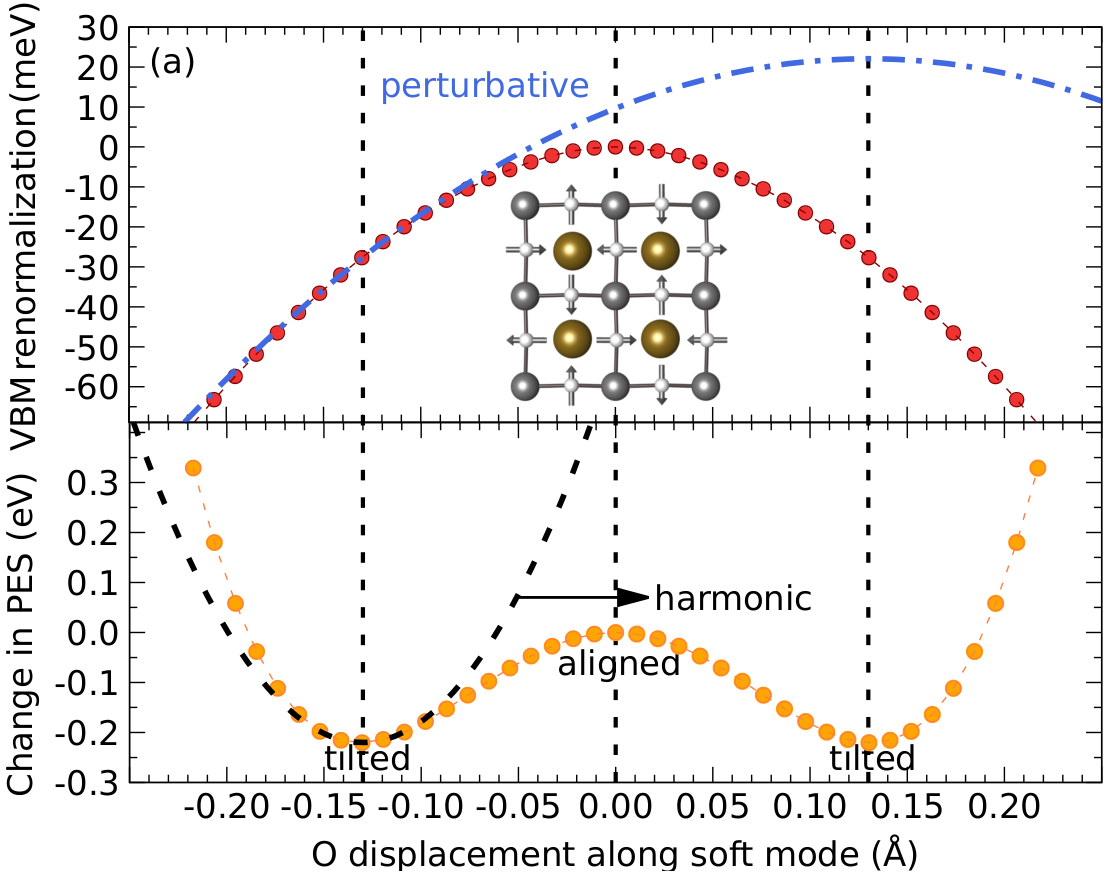}
  \hspace*{0.4cm}
   \includegraphics[width=0.42\textwidth]{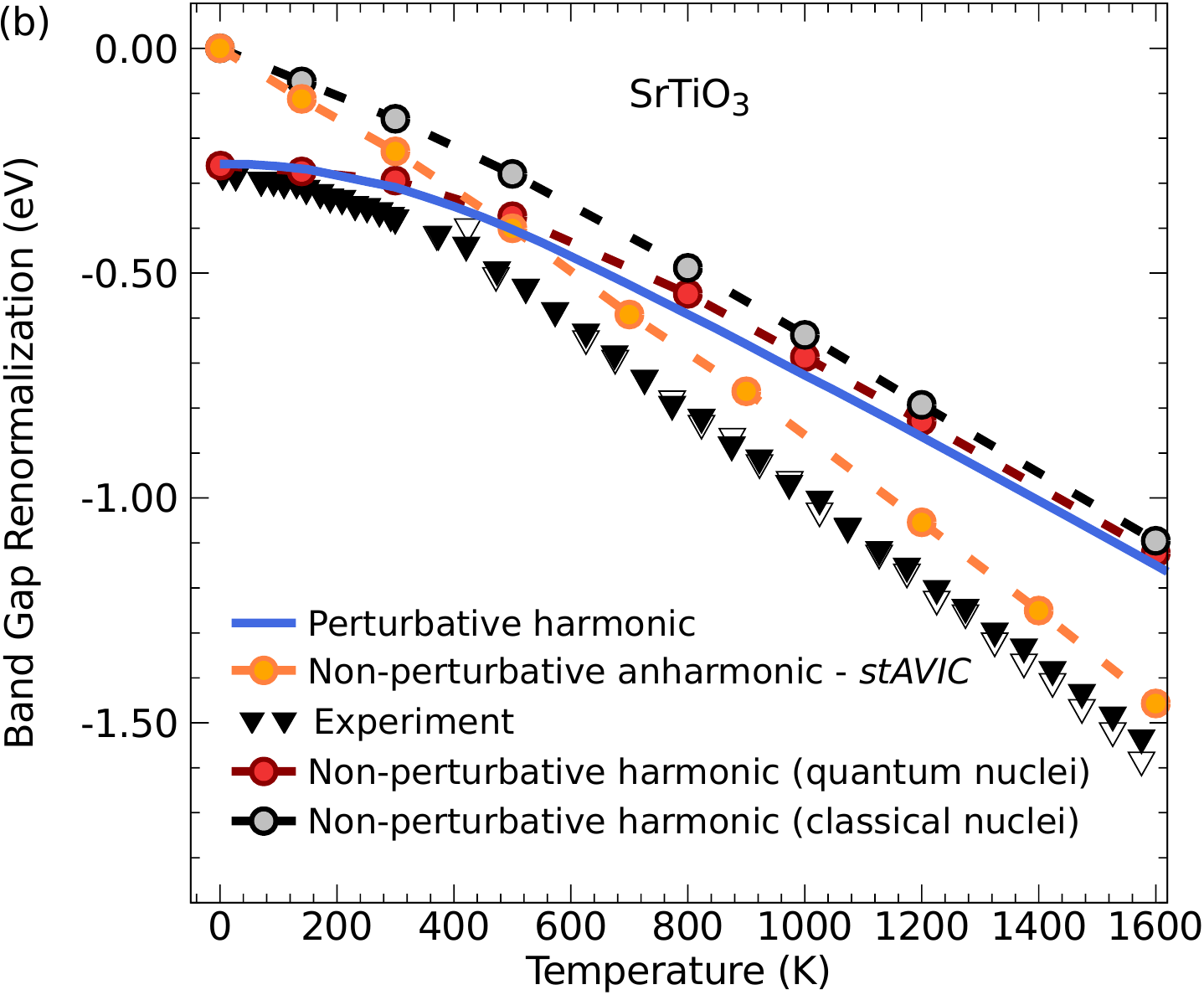}
  \caption{\label{fig4} (a) Energy of the valence band maximum~(red) and PES~(orange) of SrTiO$_3$ as function 
of the displacement of atoms along the soft phonon mode at the $R$-point. 
The direction of these displacements are shown as arrows in the planar ball-and-stick model of SrTiO$_3$. O, Ti, 
and Sr atoms are represented by white, grey, and brown spheres, respectively.
Parabolic fits at the tilted minimum are shown in blue and black. (b) Band gap renormalization of 
cubic SrTiO$_3$ as function of temperature calculated using DFT-PBE and $5\times5\times5$ supercells~(625 atoms). 
Perturbative harmonic calculations~$\Delta\<\epsilon_{\t{g}}^\t{pt}\>_T^\t{ha-qm}$ using finite 
differences~\cite{Capaz_2005} are shown in blue; non-perturbative harmonic calculations 
$\Delta\<\epsilon_{\t{g}}\>_T^\t{ha-cl}$ and $\Delta\<\epsilon_{\t{g}}\>_T^\t{ha-qm}$ in red and grey;  
   non-perturbative anharmonic \textit{stAVIC} calculations~$\Delta\<\epsilon_{\t{g}}\>_T^\t{MD}$ in orange. 
  Long-range polar interactions are accounted for in all cases, see Appendix~\ref{sec.Polar_FC}. 
  Triangles represent experimental data~\cite{Kok_2015}; the respective band gap in the 
 static limit~($3.568$~eV) was determined via linear regression~\cite{Cardona_2001} from the high $T>$~800~K data.}
\end{figure*}

\begin{figure}[t]
  \includegraphics[width=0.425\textwidth]{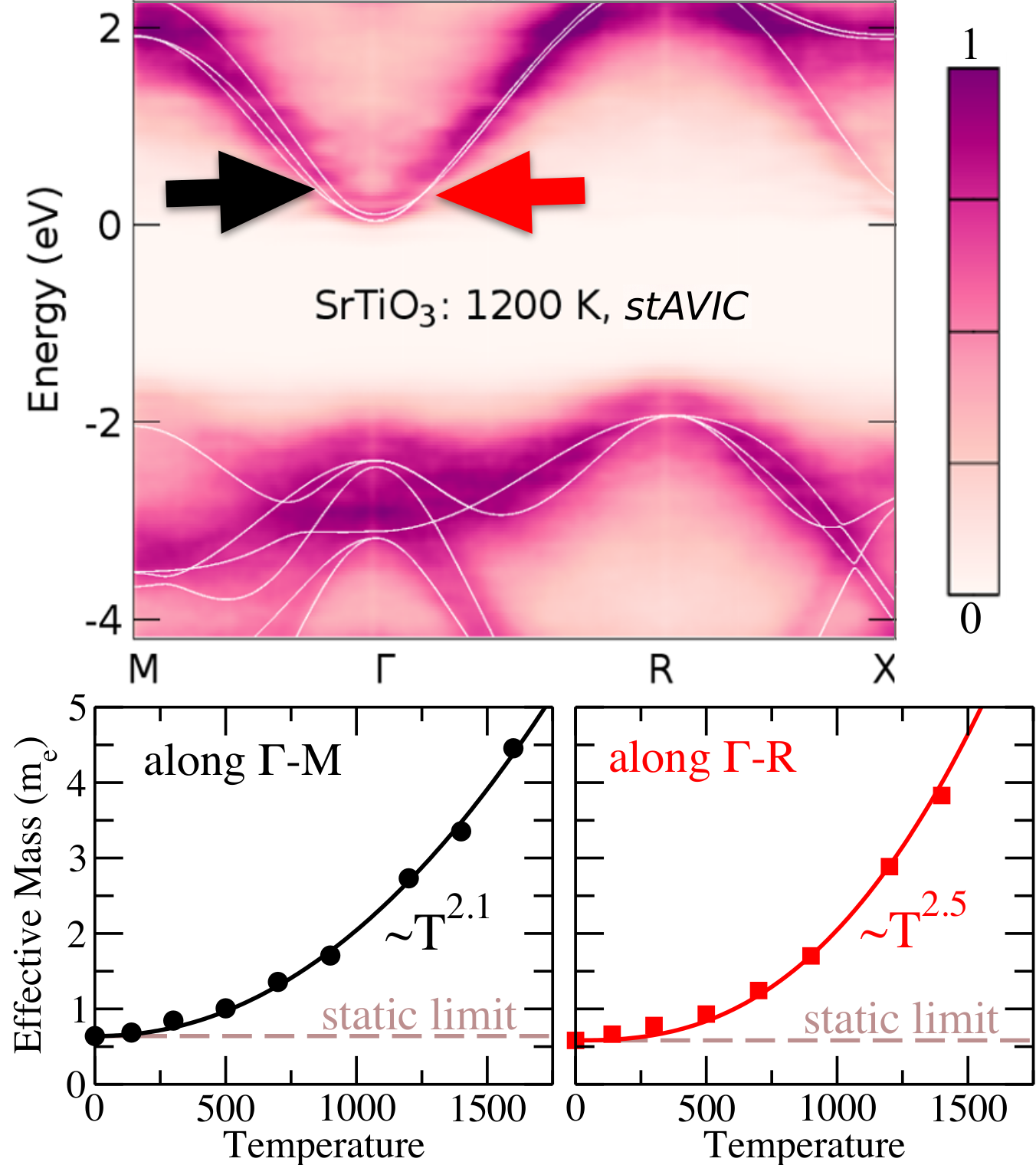}
  \caption{\label{fig5} Thermodynamically averaged spectral function $\<  A^{{\vec{R}}(t)}_{\bk} (E) \>_T$ of cubic 
SrTiO$_3$ for $T=1200$~K calculated with \textit{stAVIC} using DFT-PBE and 30 uncorrelated geometries in 
a $5\times5\times5$ supercell containing 625 atoms. The band structure in static equilibrium is shown as white 
lines. The extracted, temperature-dependent electron effective masses along $\Gamma$-R and $\Gamma$-M are shown
below.}
\end{figure}

\subsection{\textit{stAVIC}: Temperature-dependent band structure of SrTiO$_3$}

Unlike Si, AVICs are not negligible for many 
materials~\cite{Patrick_2015,Antonius_2015,Monserrat_2015_MC,Saidi_2016,Lai_2015,Tadano_2018,Asher:2020er}, 
as we demonstrate here for the prototypical perovskite SrTiO$_3$. 
At $T=0$~K, this material exhibits a tetragonal $I4/mcm$~structure~($c/a=0.998$), in which the 
individual tetrahedra are slightly tilted with respect to each other~\cite{Loetzsch_2010}. 
Above 105~K~\cite{Gogoi_2016_STO_exp} and up to its melting point at 2300~K~\cite{Kok_2015}, 
SrTiO$_3$ exhibits a cubic $Pm3m$~structure, in which all tetrahedra appear to 
be aligned, cf. Fig.~\ref{fig4}(a). This cubic structure does not correspond to a minimum, but to a saddle point 
of the PES and thus features imaginary phonon frequencies. Even in the cubic lattice~($c/a=1$), the tetrahedra favor a 
tilted arrangement  in the static limit, corresponding to the minima in Fig.~\ref{fig4}(a). 
Thermodynamic hopping between these wells results, on average, in an apparent alignment of the tetrahedra, in 
close analogy to other vibrationally-stabilized materials~\cite{Fabris:2001kx,Sternik:2005tc,Errea:2011fr,Carbogno_2014}. 
Perturbative approaches cannot capture this complex dynamics that is commonly observed in perovskites~\cite{Patrick_2015,Roekeghem_2016,Saidi_2016}: 
If the saddle point with aligned tetrahedra 
is chosen as the static equilibrium~$\vec{R}^\t{eq}$,  phonon modes with imaginary frequencies have to 
be ``frozen in''~\cite{Saidi_2016} and their coupling to the electronic-structure is neglected. 
If one of the minima with tilted tetrahedra is chosen as~$\vec{R}^\t{eq}$, both the harmonic approximation 
for the PES and the parabolic electron-phonon model become not only inaccurate, but even qualitatively 
wrong at elevated temperatures, at which multiple minima are explored, as shown by the parabolic 
fits in Fig.~\ref{fig4}(a). In other words, perturbative calculations require to 
assume either (a)~a tilted alignment at all temperatures or (b)~that the
modes responsible for the stabilization of the cubic polymorph above $105$~K are insignificant.
Neither of these assumption is justified and the breakdown of the harmonic, perturbative model has 
direct impact on the thermodynamic properties of SrTiO$_3$.

The temperature dependence of the band gap renormalization of SrTiO$_3$ is shown 
in Fig.~\ref{fig4}(b). Examples of the spectral functions calculated along $\Gamma$-R at 
$300$~K and $1200$~K are shown in Fig.~\ref{fig2}(b). 
Corrections~\cite{Nerry_Allen_2016} for long-range polar effects~\cite{Fan_1951} that are 
not fully captured within the finite aiMD supercells are included. Thermal lattice expansion 
and the associated, non-negligible band-gap opening of,~e.g.,~154 meV at 1000 K, are also 
accounted for non-perturbatively. 
Details on the treatment of lattice expansion and polar effects on the band gap renormalization     
 are given in Appendices~\ref{app.thermal_exp} and~\ref{sec.Polar_FC}, respectively. 
As discussed for Fig.~\ref{fig3}(b), the fact that $\Delta\<\ve_{\t{g}}\>_T^{\t{ha-cl}}$ and 
$\Delta\<\ve_{\t{g}}\>_T^{\t{ha-qm}}$ become comparable for $T>500$~K, implies that the use of 
classical aiMD is justified in this regime. In contrast to Si,
distinct deviations between harmonic~$\Delta\<\ve_{\t{g}}\>_T^{\t{ha-cl}}$ and 
anharmonic~$\Delta\langle\ve_{\t{g}}\rangle_T^\t{MD}$ data are observed for SrTiO$_3$, leading to an 
additional renormalization in \textit{stAVIC} as large as 147~meV at 600~K and 260~meV at 1200~K. 
With respect to the perturbative, harmonic data, this corresponds to a remarkable increase 
of 18~\% and 27~\%, respectively. With respect to experiment~\cite{Kok_2015}, \textit{stAVIC} 
improves the agreement significantly and quantitatively reproduces the measured high-temperature slope.
This has substantial influence on the actual properties of SrTiO$_3$,~e.g.,~this band-gap narrowing
massively increases intrinsic charge carrier densities~$n_c
\propto \exp\left[-\<\epsilon_{\t{g}}\>_T/(2k_{\rm B}T)\right]$  by two orders of magnitude 
at 1000~K.
% due to the band-gap's temperature dependence, a large fraction of which is driven by AViCs.

More insights can be obtained from the momentum-resolved spectral-functions,~e.g.,~by extracting
the electron effective masses~$m^*_e(T)$ along $\Gamma$-R and $\Gamma$-M, as done in Fig.~\ref{fig5}. 
These particular effective masses have been topic 
of debate~\cite{Mechelen_2008,Allen_S_2103}, 
since \textit{ab initio} calculations~\cite{Marques_2003,Janotti_2011} of SrTiO$_3$ systematically underestimate 
measured values by a factor of two or more~\cite{Janotti_2011,Ahrens:2007hk}. 
The \textit{stAVIC} calculations reveal a large enhancement of~$m^*_e(T)$ with $T$, confirming the important 
role of the nuclear motion suggested by experiments~\cite{Mechelen_2008,Allen_S_2103}.
The strong temperature dependence $\propto T^{2.1-2.5}$ also
substantiates the hypothesis~\cite{Frederikse:1964kj} that~$m^*_e(T)$
is responsible for the unusually large decrease in Hall mobility~$\propto T^{-2.7}$ at high-temperatures~\cite{Tufte_1967}, 
that defies harmonic models yielding~$\propto T^{-1.5}$~\cite{Frederikse:1964kj}. 
Certainly, this dictates further research along these lines,
 so as to disentangle the influence of AVICs on, e.g., charge carrier densities, effective masses, 
scattering mechanisms~(lifetimes and linewidths), as well as on the interplay 
with polaronic~\cite{Sio:2019im} and quantum-nuclear effects at low~$T$~\cite{Ramirez:2006bg}.

\section{Conclusions} \label{Sec.Conclusions}

In this work, we have demonstrated a fully anharmonic, non-perturbative theory of 
the vibronic interactions in solids that overcomes the two main 
approximations~(harmonic and electron-phonon coupling model) that limit the applicability of 
perturbative state-of-the-art formalisms~\cite{Antonius_2015,Monserrat:2015jj,Saidi_2016}. 
The presented \textit{stAVIC} methodology gives access to momentum-resolved electronic spectral
functions and, in turn, to a plethora of other electronic properties~\cite{FG_review}. As demonstrated for the 
perovskite  SrTiO$_3$, accounting for AVICs is pivotal at elevated temperatures and/or in strongly anharmonic
materials. \textit{stAVIC} thus lends itself to aid and guide the in-silico materials design 
for high-temperature applications,~e.g.,~for optical gas sensing in next-generation combustion 
chambers~\cite{Schultz:2015hi}, solid-oxide fuel cells~\cite{Marina:2002kka,Skinner:2001fv}, thermoelectric waste-heat 
recovery devices~\cite{Ohta:2007fua,Fergus:2012jha}, as well as hybrid photovoltaic cells 
operating under concentrated sunlight~\cite{Brunauer:2016jv}. For all these applications, in which perovskites, but also many other 
highly-anharmonic materials, play a substantial role, an accurate assessment of the temperature-dependent, momentum-resolved electronic structure is essential, since the associated electronic properties 
such as band gaps and effective masses~\cite{Huo:2018cs,Pei:2012ih}, as well as the anisotropic 
band-structure corrugation~\cite{Chen:2013cn} are critical for the material's performance.

\begin{acknowledgments}
CC thanks Hagen-Henrik Kowalski, Florian Knoop, and Friedhelm Bechstedt for fruitful discussions.
This project was supported by TEC1p (the European Research Council (ERC) Horizon 2020 
research and innovation programme, grant agreement No. 740233), 
BigMax (the Max Planck Society’s Research Network on Big-Data-Driven Materials-Science), 
and the NOMAD pillar of the FAIR-DI e.V. association. All the electronic-structure theory 
calculations produced in this project are available 
on the NOMAD repository:  http://dx.doi.org/10.17172/NOMAD/2020.03.18-1.
\end{acknowledgments}

\appendix
\numberwithin{figure}{section}

\section{Thermal lattice expansion of SrTiO$_3$} \label{app.thermal_exp}

\begin{figure}[hbt!] 
  \includegraphics[width=0.425\textwidth]{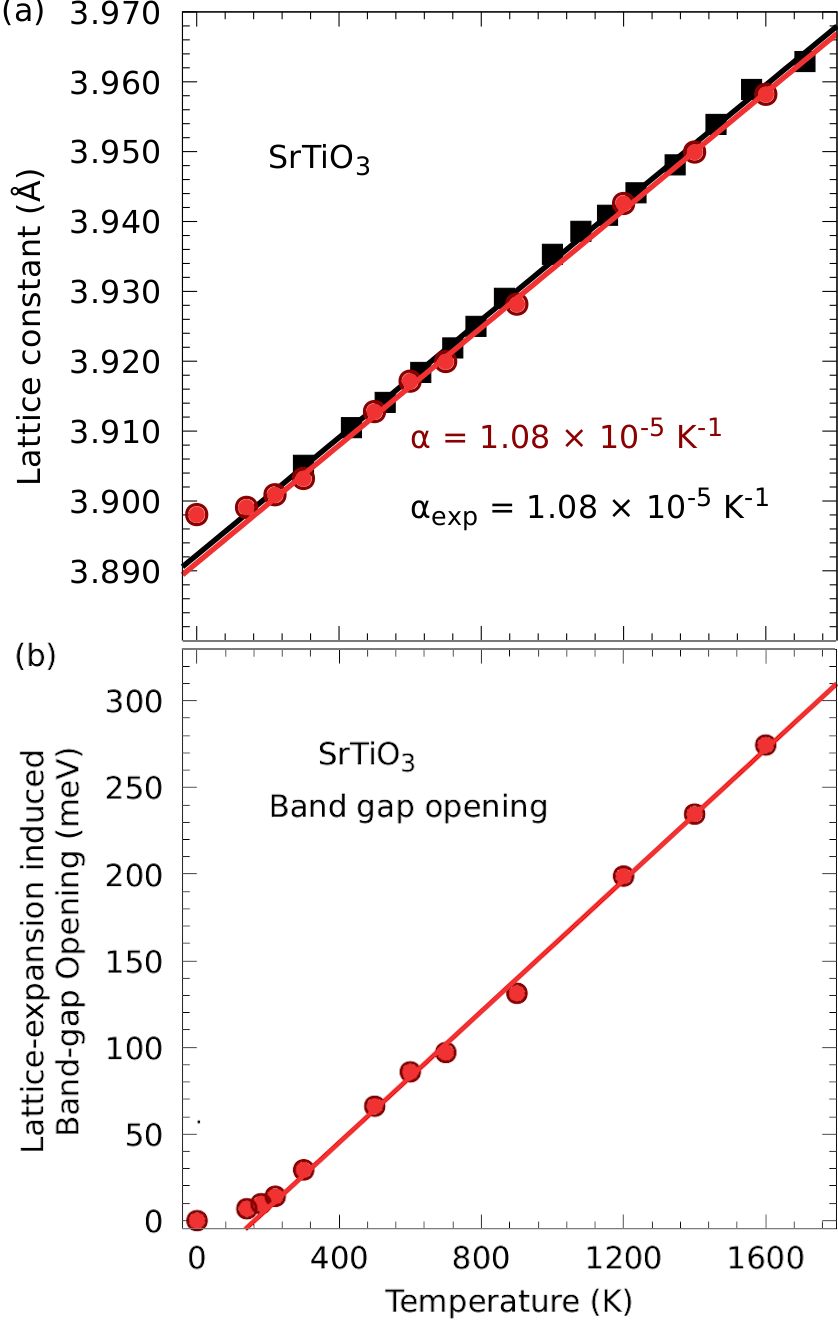}
  \caption{ 
  \label{A1}
  (a) Calculated lattice constant of cubic SrTiO$_3$ as a function of temperature (red discs) 
  versus experimental data from Ref.~[\onlinecite{Ligny_1996}] (black squares). 
  For each case the linear thermal expansion coefficient is indicated. 
  The linear fits for $T>300$~K are guides to the eye. 
  (b) Temperature-dependent renormalization of the band gap of cubic 
  SrTiO$_3$ due to thermal lattice expansion. The linear fit 
  for $T>300$~K is a guide to the eye.}
\end{figure}

The computed temperature dependence of the lattice constant of 
SrTiO$_3$, which is associated to its thermal expansion,
was calculated by running aiMD in 6$\times$6$\times$6 supercells containing 1080 atoms, 
thus taking anharmonic effects into account~\cite{Roekeghem_2016}. As shown in Fig.~\ref{A1}(a), we observe
a linear increase of the lattice constant for $T>200~K$, i.e., for 
temperatures in which the cubic structure is indeed stable. 
This corresponds to a considerable linear thermal expansion coefficient 
of~$\alpha_{\rm L} = \frac{1}{a(T)} \frac{\partial a(T)}{\partial T} \approx 1.1 \times 10^{-5}$~K$^{-1}$, 
which is in excellent agreement with the corresponding experimental 
value extracted from the measured data reported in Ref.~[\onlinecite{Ligny_1996}]. The lattice 
expansion induces a significant opening of the band gap as the temperature 
increases, for example 154~meV at 1000~K, as shown in Fig.~\ref{A1}(b). 
This effect is accounted for in all calculations of the band gap renormalization of
SrTiO$_3$ discussed in the main text. Let us note that 
thermal lattice expansion has not been accounted for 
in the {\it stAVIC} calculations for the band gap renormalization of Si
to allow for a consistent comparison to literature data. Furthermore, 
our calculations reveal that this effect is negligible in Si even at high temperatures,
resulting, for example, to a band gap opening of 13~meV at $1100$~K.  
%, in which these effects are not considered as well.

%\CC{Note: The lattice expansion is defined as $\alpha = \frac{1}{a} \frac{\partial a(T)}{\partial T}$, with $a$ the lattice constant. \MZ{ The quantity you wrote is the linear coefficient of thermal expansion.}
%You are showing the temperature dependence of the lattice constant. Please adapt the labels.}

\begin{figure}[hbt!] 
  \includegraphics[width=0.49\textwidth]{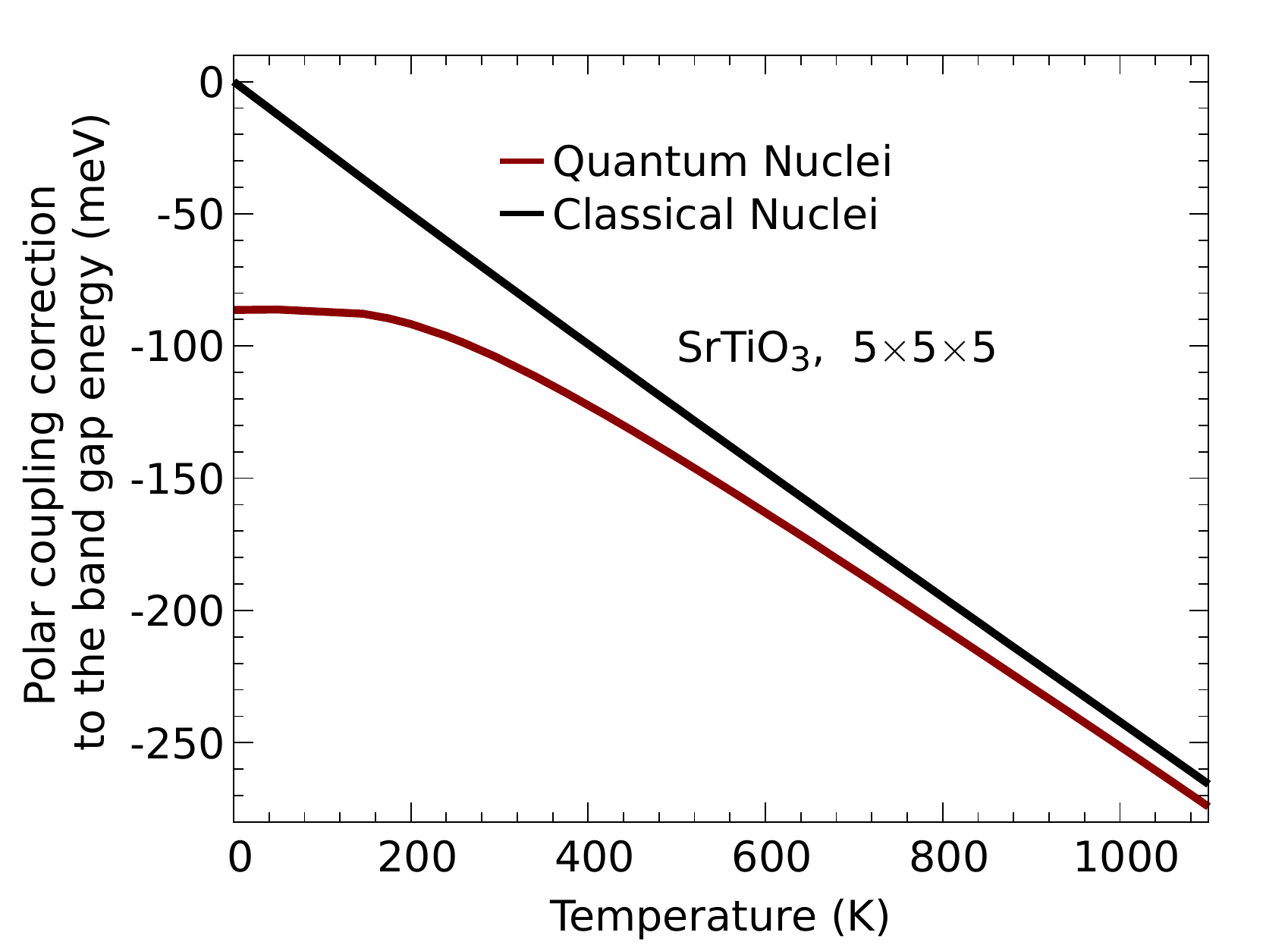}
  \caption{ 
  \label{C1}
Fr\"ohlich polar coupling correction [Eq.~\eqref{eq.Fr_contr}] to the band gap of 
SrTiO$_3$ for a 5$\times$5$\times$5 supercell as a function of temperature using quantum~(red) and classical~(black) occupation numbers.}
  \end{figure}

\section{ Polar Fr\"ohlich Coupling} \label{sec.Polar_FC}
In polar materials, there is an additional contribution to the energy level renormalization stemming from 
long-range Fr\"ohlich coupling~\cite{Fan_1951,Verdi_2015,Sjakste_2015,Nerry_Allen_2016} that is not fully 
captured in the limited supercells used in aiMD. We account for the missing portion of these effects via 
the following analytic correction for the adiabatic case~\cite{Fan_1951,Nerry_Allen_2016,Zacharias_2020}:
\begin{equation}\label{eq.Fr_contr}
\<\Delta\ve_{l}^{\rm Fr}\>^{\rm HA}_T =  \frac{2}{\pi} \alpha \hbar \w_{\rm LO} 
\, {\rm tan}^{-1}\bigg(\frac{q_{\rm F}}{q_{\rm LO}}\bigg) \, [2n_{T}+1]\;,
\end{equation}
which can be obtained by integrating Eq.~(B3) of Ref.~[\onlinecite{Zacharias_2020}] up 
to the truncation parameter $q_{\rm F}$ defined below. 
In Eq.~\eqref{eq.Fr_contr}, the strength of the polar coupling 
is characterised by the
dimensionless polaron constant given by~\cite{Mahan}:
\begin{equation}\label{eq.Fr_contr_2}
\alpha = \frac{e^2}{4 \pi\e_0 }\frac{1}{\hbar} \bigg( \frac{1}
{\kappa_{\infty}}-\frac{1}{\kappa_0}\bigg) \bigg(\frac{m^*}{2\hbar\w_{\rm LO}} \bigg)^{1/2}
\end{equation}
 where $m^*$ is the effective mass of the carrier, $\w_{\rm LO}$ is the 
frequency of the LO phonon, and $\kappa_{\infty}$,
$\kappa_{0}$ are the high-frequency dielectric constant and static permittivity, respectively.
The quantity $n_{T}  = [\exp(\hbar\w_{\rm LO}/k_{\rm B}T)\!-\!1]^{-1}$ is the Bose-Einstein occupation factor
of the LO mode, and $q_{\rm LO}$ is defined as $\sqrt{2m^*(\w_{\rm LO}+\w)/\hbar}$, where $\hbar \omega$ is the energy of the state. %For calculating the contribution of the Fr\"ohlich coupling to the band gap renormalization we set the energy of the valence band maximum at zero, and the energy of the conduction band bottom at 3.45~eV. This latter value is equivalent to the band gap calculated using the HSE06-vdW functional. 
The radius of integration $q_{\rm F}$ is used as a truncation parameter
to avoid a double counting of the Fr\"ohlich interactions already accounted 
for in the aiMD supercell. For our calculations we take $q_{\rm F}$ equal to the radius of the Debye sphere,
i.e., to the sphere with the same volume as the Brillouin zone of the employed 5$\times$5$\times$5 supercell. 
In order to evaluate Eq.~\eqref{eq.Fr_contr} we set the LO phonon energy to 
$\hbar \w_{\rm LO} = 59$~meV~\cite{Servoin_1980}, the light-electron and light-hole effective 
masses along R$\rightarrow \Gamma$ to $m_e = 0.537 $ and $m_h = 0.944$ ~\cite{Marques_2003}, 
and the high-frequency dielectric constant to $\kappa_{\infty} = 5.5$~\cite{Cowley_1964}. 
The static permittivity of SrTiO$_3$ exhibits a large variation with temperature, and therefore 
we extract the temperature dependence of $\kappa_{0}$ from Ref.~[\onlinecite{Servoin_1980}]. Our calculations of the temperature 
dependent polar coupling correction to the band gap of SrTiO$_3$ for a 5$\times$5$\times$5 supercell are shown in Fig.~\ref{C1}.
For classical nuclei (black), the square bracket in Eq.~(\ref{eq.Fr_contr}) is replaced by its classical limit without zero-point
vibrations, which results in a correction that varies linearly from 0~meV at $T=0$~K to -265~meV at $T=1100$~K.

\section{Evaluation of band gaps via the joint density of states} \label{app.JDOS}

\begin{figure}[hbt!]
  \includegraphics[width=0.44\textwidth]{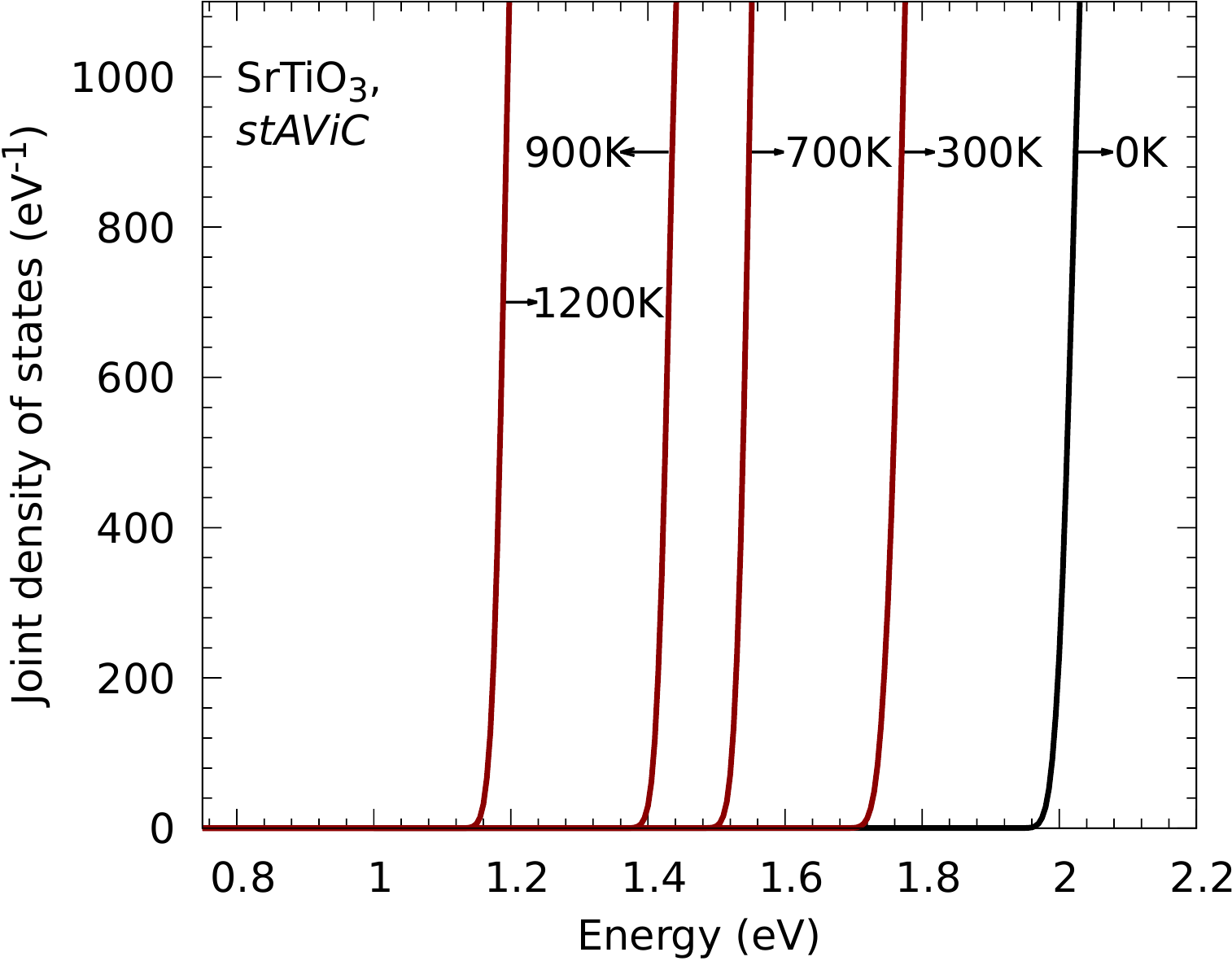}  
\caption{
\label{B1}
Temperature-dependent joint density of states of SrTiO$_3$ calculated using {\it stAVIC} 
and 5$\times$5$\times$5 supercells containing 625 atoms. 
The black line represents the joint density of states evaluated with atoms at static equilibrium.} 
  \end{figure}

For the calculation of the joint density of states at each aiMD step~$\vec{R}(t)$ we consider the following relationship: 
\begin{equation}
    J^{\vec{R}(t)}(E) = \sum_{c\bk_c,u\bk_u} \delta\Big(\epsilon^{\vec{R}(t)}_{c\bk_c} - 
\epsilon^{\vec{R}(t)}_{u\bk_u} - E \Big)
\end{equation}
where the summation runs over all conduction and valence states indices ${c\bk_c}$ and $u\bk_u$. 
In order to determine the temperature dependence of the band gap we consider the energy offset 
between the thermodynamically averaged joint density of states for 
temperatures $T$ and $0$~K, i.e., the energy offset between 
$ \< J^{\vec{R}(t)}(E) \>_T$ [red lines in Fig.~\ref{B1}]  
and  $\< J^{\vec{R}(t)}(E) \>_{0 \rm K}$ [black line in Fig.~\ref{B1}]. 

\newpage

\bibliography{references}{} 
\end{document}